# Using Logs to support Programming Education

Gilmar Gomes do Nascimento[1,2] [0009-0002-1319-9824] , Maria Claudia F.P Emer[2][0000-0002-0963-1891] , Adolfo Gustavo Serra Seca Neto[2][0000-0002-0260-5922] and Laudelino Cordeiro Bastos[2][0000-0002-1585-7168]

[1] Instituto Federal de Educação, Ciência e Tecnologia do Amazonas, Boca do Acre AM, BRA
[2] Universidade Tecnológica Federal do Paraná, Curitiba PR, BRA

gilmar.nascimento@ifam.edu.br, adolfo, bastos{@utfpr.edu.br}, mciemer@gmail.com

**Abstract.** Software developers use metrics to evaluate code quality and productivity, but these practices are still rare in programming education. This project bridges the gap by collecting real-time learning analytics from individual student and whole-class code development logs. This granular, quantitative data provides educators with qualitative insights into the learning process. It allows them to evaluate student comprehension, identify common challenges, and critically assess whether the allocated time for exercises and algorithms is sufficient for mastery. Unlike traditional Learning Management Systems, we propose a novel approach: a plugin for a widely used code editor that captures granular interactions during programming and documentation. The resulting dataset logs coding behaviors, errors, and progress, enabling evidence-based analysis of learning patterns and educational benchmarking. By structuring this real-time programming trail, we support research on teaching methodologies, learner challenges, and skill acquisition. Quantitative metrics complement qualitative assessment by evaluating code, exercise progress, and timestamp logs. Our goal is to provide an open-access database for educators and researchers, fostering data-driven insights to enhance instruction and personalize learning experiences. This work aligns industrial best practices with pedagogical innovation, advancing measurable, empirical approaches to programming education.

## 1 Introduction

While traditional teaching focuses on reading and writing text, programming courses introduce a unique language: algorithms—logical sequences interpretable by both humans and machines. This subject, which is part of technology courses, is “considered challenging for students and teachers” [1], as it requires the development of formal logic.

During code development, tools generate automatic logs, capturing everything from syntax errors to debugging patterns. This data has the potential to transform programming education, but its application faces critical challenges: existing tools (such as online judges or Learning Management Systems - LMS) are limited to analyzing final submissions, overlooking the learning process itself [2].

We propose a system that captures incremental code development. Our system captures real-time logs through a plugin integrated with popular code editors (e.g., VS Code). Unlike platforms like Moodle—, which only record generic interactions (material access, forum participation) [3]—our system tracks the dynamic evolution of code, enabling:

- Identifying learning patterns [4], and
- Personalizing instruction and detecting plagiarism [5].

The relevance of this proposal lies in its direct applicability to technical and higher education institutions. By transforming logs into actionable feedback, instructors can intervene proactively, while students gain autonomy to correct deviations in their learning process. Additionally, the generated dataset—unprecedented in its granularity—opens new avenues for research in computational education, providing empirical evidence on effective pedagogical strategies [6].

Teaching programming is a multidisciplinary activity that requires refined pedagogical practice—that is, constant reflection that integrates theory and practice in the classroom. A common strategy is to use everyday examples to encourage problem solving through algorithms. However, evaluating the quality of these algorithms requires going beyond functional verification.

These algorithms, defined as computational procedures that transform inputs into outputs, generate records known as logs. But how can these logs be used to transform teaching and learning? This text explores the intersection between technology and education, showing how log analysis can offer valuable insights for personalizing teaching and identifying student difficulties. These interactions, in addition to transforming inputs into outputs, generate logs. According to Sah (2002, apud [7]), "a log is defined as a set of time-stamped records, which only supports insertion, and which represents events that have occurred on a computer or network device." Furthermore, according to [7], “a log usually contains a large number of entries, also known as log messages.” In general, the log level, its content, and the appropriate location to declare it are decisions made by developers during software creation.

## 2 Related Work

Log storage and analysis constitute a well-established research area in software engineering, with applications in telemetry and system monitoring. However, in the educational domain - particularly in programming instruction - their potential remains under explored. While professionals commonly use these techniques to track assignment submissions and platform access, recent research demonstrates their ability to uncover fundamental aspects of the learning process.

Research on the use of logs in programming education seems to constitute a niche area of research that is still limited, mainly explored by teachers and researchers who seek to deepen and formalize specific pedagogical practices. The scarcity of course completion papers in the area corroborates this characteristic. A search on the SBC portal using the terms: log OR logging AND “teaching of programming” returned few or no results, indicating that the topic is still in its infancy in national undergraduate academic production.

After conducting a systematic review of the literature, we identified relevant studies and synthesized their key characteristics in Table 1, highlighting similarities to our approach.

**Table 1.** Studies on Programming and learning Environments. **Source**: The authors, 2025.

| Environment | Title | Country |
|---|---|---|
| IDE/Online | Analysis of C Programming Performance: A correlational study of novice programmers compilers error log | Philippines |
| OJ/IDE | Using learning analytics in the Amazonas: understanding students’ behavior in introductory programming | Brazil |
| IDE/Online/OJ | Previsão de indicadores de dificuldade de questões de programação a partir de métricas do código de solução | Brazil |
| IDE | Introducing Thonny, a Python IDE for Learning Programming | Estonia |

This study conducts a systematic review of existing approaches, focusing specifically on:

1. Techniques for collecting and processing educational logs;
2. Pedagogical analysis methods capable of identifying learning patterns, persistent difficulties, and problem-solving strategies adopted by students.

Despite its consolidated use in software development tools, the application of telemetry in programming teaching is not yet a widely used practice. Its potential, however, is significant: institutions can use telemetry data to identify specific difficulties students have in understanding the material, assess the level of challenge presented by their courses, and, consequently, intervene more precisely and personalized.

Given this potential, this research aims to understand the process of building algorithms and the software engineering tools and techniques used in teaching, through logs created and stored during the teaching-learning process at a Federal Institute in the Amazon region. With this collection, we will have qualitative and quantitative information on code production, which will be fundamental for formulating models with potential applications in teaching programming and developing solutions.

Teaching programming is not limited to the transmission of technical knowledge, but also covers aspects such as code licensing, good development practices, and academic integrity. Understanding the evidence/trace left in the production of code can reveal signs of plagiarism, as the coupling of code snippets with matching timestamps and the standardized use of reserved words signal possible fraud, improper copying, and/or repetition of code. In this context, log storage emerges as a valuable resource to support student assessment.

For a clear and unified understanding of the elements that make up the activity of logging, we adopt the taxonomy proposed by [8], which systematically defines the main concepts:

- Log statement
- Log message
- Log
- Log Level
- Log Content
- Log location
- Log placement
- Logging
- Logging practice

The authors of [9] confirmed that teams use logs primarily to analyze software problems. They also observed other uses, such as in test development and requirements reverse engineering. The authors identified 13 types of information that developers seek in logs, the most frequent being: error propagation between systems, timestamps associated with log lines, data flow, interaction between software components, and differences between multiple executions.

## 3 Methodology

This study investigates algorithmic learning processes in Amazonian technical education through IDE interaction logs, combining software engineering metrics with pedagogical analysis. It aims to develop and evaluate a framework for improving programming education through telemetry and learning analytics. To achieve this, a design-oriented approach is required.

To guide the formulation of models and applied frameworks for programming education, this research adopts the Design Science Research (DSR) methodology. With its roots in engineering and the sciences of the artificial [10], DSR provides a rigorous paradigm for creating and evaluating artifacts—such as models, methods, and constructs—intended to solve identified organizational problems and improve effectiveness.

This approach is particularly relevant for Information Systems (IS) research, where the community often measures research relevance by its practical applicability. As Hevner [11] emphasizes, DSR directly addresses this need by grounding research outcomes in real-world utility. Following this rationale, our work aligns with the established DSR process model proposed by Peffers et al. [10]. While the model comprises six core activities, the present study focuses specifically on the following four:

- Problem Identification and Motivation
- Definition of Objectives for a solution
- Design and Development
- Demonstration

## 4 Expected Outcomes and Impact

This article investigates the use of interaction logs within programming education, with a specific focus on Brazil's Federal Institutes of Education, Science, and Technology.

We conducted a systematic literature review to identify existing applications and research concerning educational logs. Our review revealed a significant diversity of programming tools and environments across institutions, which presents a challenge for consistent data collection. To address this challenge, we propose the development of an extension for an open-source code editor. This plugin standardizes the collection of development logs in academic settings, filling a critical gap in current IDE capabilities. The subsequent phases of this project will test and implement the plugin within academic environments to refine its functionality, followed by widespread dissemination to other institutions.

While the collected logs may initially lack standardization and may not directly measure performance or learning, their aggregation into a centralized database is a primary goal. This database would serve as a foundation for this research and enable future applications, such as large-scale learning analytics (big data in education), quantitative pedagogical assessment, and the identification of common student learning obstacles.